\documentclass{revtex4} %twocolumn,[preprint]
\usepackage{mathrsfs}
\usepackage{bm}
\usepackage{amsmath}
\begin{document}
\title{Two-dimensional Inside-out Eaton Lens: Wave Properties and Design}
\author{Yong Zeng and Douglas H. Werner}
\affiliation{Department of Electrical Engineering, Pennsylvania
State University, University Park, PA 16802}
\input epsf
\begin{abstract}
In this paper we study two-dimensional inside-out Eaton lens
theoretically and numerically. With the help of the WKB
approximation, we investigate the finite-wavelength effect
analytically and demonstrate one necessary condition for perfect
imaging by the lens. Furthermore, we present one example design
and test its performance by using full wave Maxwell solvers.
\end{abstract}
 \maketitle

More than half a century ago, it was proposed that gradient index
lenses \cite{grin}, such as the Maxwell fish-eye lens
\cite{maxwell}, the Luneburg lens \cite{luneburg} and the Eaton
lens \cite{Eaton}, can be free of geometrical aberrations and form
perfect images, at least at the geometrical-optics level (see
Reference \cite{leonhardt1} and \cite{leonhardt2} for more
details). Taking the inside-out Eaton lens as an example, its
refractive index $n(r)$ equals $\sqrt{(2-r)/r}$ for $1\leq r\leq
2$ and 1 otherwise, with $r$ being the radius (see Figure 1). We
can analytically prove that light rays emitted from a source at
position $\mathbf{r_{0}}$, with $r_{0}<1$, will be focused exactly
at position $-\mathbf{r_{0}}$ \cite{leonhardt2}. Therefore, like
Maxwell's fish-eye lens, perfect imaging can be obtained by an
inside-out Eaton lens, and both the source and the image are
inside the optical instruments.

Recently, a renaissance of scientific interests appears in these
gradient index lenses
\cite{ma,Kundtz,Smolyaninova,smith1,Zentgraf,leonhardt3},
partially because of the developments of metamaterials
\cite{pendry1,solymar} and transformation optics
\cite{pendry2,leonhardt4,kwon,chen}. Metamaterials are manmade
media whose effective permittivities and permeabilities are
determined by their deeply subwavelength structures as well as
their constituent materials \cite{solymar}. For instance,
integrating split ring resonators with metallic rods will result
in a metamaterial with an effectively negative index of refraction
\cite{smith}. Transformation optics, on the other hand, is
inspired by an intriguing property of Maxwell's equations,
\textit{i.e.}, their form is invariant under arbitrary coordinate
transformations, assuming the field quantities and the material
properties are transformed accordingly \cite{pendry2,leonhardt4}.

In this paper, we design a two-dimensional dielectric inside-out
Eaton lens consisting of metallic wires in a homogeneous
background medium with positive permittivity. To investigate the
effect of the finite wavelength, we employ the WKB approach to
solve the wave equations \cite{chew,sakurai}, and obtain one
necessary condition for perfect imaging. Finally we test our
design with full wave Maxwell solvers.

\begin{figure}
\epsfxsize=250pt \epsfbox{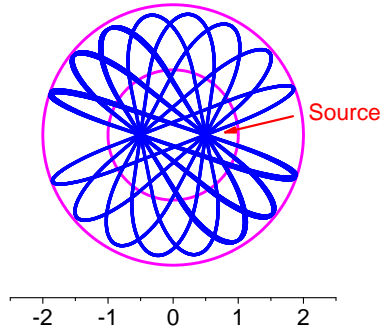} \vspace*{-7.0cm}
\caption{Inside-out Eaton Lens ($1\leq r\leq2$). The source is
located at $r_{0}=0.5$. Light rays (blue) are described by
Hamilton's equation (A.2).} \label{fig1}
\end{figure}

The original Eaton lens has a spherical geometry and an index of
refraction as
\begin{equation}
n(r)=\sqrt{\frac{2}{r}-1}, \label{eq1}
\end{equation}
with $r$ being the radius \cite{Eaton}. Since it possesses
spherically symmetry, the entire ray trajectory lies in a plane
which is orthogonal to a conserved angular momentum $\mathbf{L}$
\cite{leonhardt2} (See Appendix A). Consequently, we can simplify
it to be a two-dimensional cylinder with similar refractive index.
We further assume that the cylinder lies in the $xy$ plane, as
well as the propagation plane of the light ray. Moreover, we only
consider TM-polarized light where the magnetic-field vector
$\mathbf{H}$ points in the $z$ direction. By sacrificing the
impedance matching, we can further assume that the two-dimensional
cylinder is purely electrical, \textit{i.e.} $\mu=1$, with its
permittivity $\epsilon(r)$ given by $n^{2}(r)$.

A detailed ray-optics description of the inside-out Eaton lens can
be found in Ref. \cite{leonhardt2}, and it is repeated in Appendix
A for the convenience of readers. We therefore only consider the
wave interpretation here. Starting from the following wave
equation
\begin{equation}
\frac{1}{\epsilon(r)}\nabla^{2}H+\nabla\frac{1}{\epsilon(r)}\cdot\nabla
H=-\frac{\omega^{2}}{c^{2}}H, \label{eq2}
\end{equation}
with
$\nabla=\partial_{x}\mathbf{e}_{x}+\partial_{y}\mathbf{e}_{y}$ and
$H$ as $H_{z}$, this equation can be reformulated in cylindrical
coordinates as
\begin{equation}
\frac{\partial^{2}H}{\partial
r^{2}}+\left(\frac{1}{r}-\frac{1}{\epsilon}\frac{d\epsilon}{dr}\right)\frac{\partial
H}{\partial r}+\frac{1}{r^{2}}\frac{\partial^{2}H}{\partial
\theta^{2}}+k_{0}^{2}\epsilon H=0, \label{eq3}
\end{equation}
with $k_{0}=\omega/c$ being the wave number in free space. We now
assume that the magnetic field $H(r,\theta)$ can be expanded as
$\sum_{n}f_{n}(r)e^{in\theta}\sqrt{\epsilon/r}$, where the
functions $f_{n}$ satisfy
\begin{equation}
f_{n}''+\left(k_{0}^{2}\epsilon-\frac{n^{2}}{r^{2}}+\frac{\epsilon^{2}-3r^{2}\epsilon'^{2}+2r\epsilon\epsilon'+2r^{2}\epsilon\epsilon''}{4r^{2}\epsilon^{2}}\right)f_{n}=0.
\label{eq4}
\end{equation}
In the region $r\leq1$ where $\epsilon(r)=1$, we have
\begin{equation}
f_{n}''+\left(k_{0}^{2}-\frac{n^{2}-1/4}{r^{2}}\right)f_{n}=0,
\label{eq5}
\end{equation}
and express the solutions generally as
\begin{equation}
f_{n}(r)=\sqrt{r}\left[a_{n}H^{(1)}_{n}(k_{0}r)+b_{n}H^{(2)}_{n}(k_{0}r)\right],
\label{eq6}
\end{equation}
where $H^{(1)}_{n}$ and $H^{(2)}_{n}$ are the $n$-th order Hankel
functions of the first and second kind, respectively. In the
region $1<r<2$, we can rewrite Equation (\ref{eq4}) as
\begin{equation}
f_{n}''+s(r)k_{0}^{2}f_{n}=0, \label{eq7}
\end{equation}
with
\begin{equation}
s(r)=\frac{2-r}{r}-\frac{n^{2}-1/4}{r^{2}k_{0}^{2}}-\frac{r+1}{r^{2}(2-r)^{2}k_{0}^{2}}.
\label{eq8}
\end{equation}
A few examples are plotted in Figure (2). For a modest $n$, $s(r)$
generally monotonically decreases from a positive value to
negative infinity. It is however always negative when $n$ is large
enough.

\begin{figure}
\epsfxsize=250pt \epsfbox{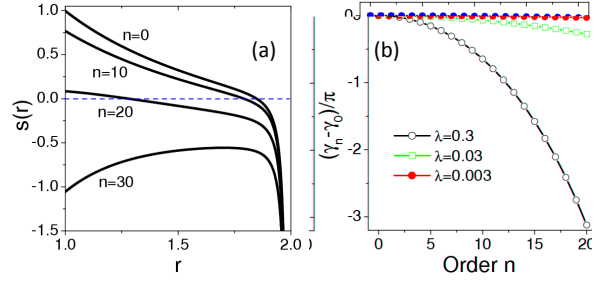} \vspace*{-7.5cm}
\caption{(a) Dependence of the function $s(r)$ on the mode order
$n$. Here the wavelength is $0.3$. (b) Effect of the wavelength
$\lambda$ on the phase factor $\gamma_{n}$.} \label{fig2}
\end{figure}

We employ the WKB approximation, developed by Wentzel, Kramers and
Brillouin in 1926, to analytically solve Equation (\ref{eq7}) with
a modest $n$ \cite{chew,sakurai}. More specifically, we assume
that $f_{n}(r)$ has the form $A_{n}e^{ik_{0}\tau(r)}$ for positive
$s(r)$, and $\tau(r)$ can be further expanded in terms of $k_{0}$,
\begin{equation}
\tau(r)=\tau_{0}(r)+\frac{1}{k_{0}}\tau_{1}(r)+\frac{1}{k^{2}_{0}}\tau_{2}(r)+\cdots.
\label{eq9}
\end{equation}
Similar arguments also hold by $\tau'(r)$ as well as $\tau''(r)$.
By collecting the leading-order terms, it is found that
\begin{equation}
\left(\frac{d\tau_{0}}{dr}\right)^{2}=s(r),\:\:\frac{d\tau_{1}}{dr}=\frac{i\tau_{0}''}{2\tau_{0}'}.
\label{eq10}
\end{equation}
Consequently the first order solution can be expressed as
\begin{equation}
f_{n}(r)\sim\frac{A^{+}_{n}}{s(r)^{1/4}}\exp\left[ik_{0}\int_{r_{n}}^{r}\sqrt{s(r')}dr'\right]+\frac{A^{-}_{n}}{s(r)^{1/4}}\exp\left[-ik_{0}\int_{r_{n}}^{r}\sqrt{s(r')}dr'\right],
\label{eq11}
\end{equation}
with $r_{n}$ standing for the turning point where $s(r)=0$. The
first term on the right hand side corresponds to an out-going wave
because its phase increases with distance, while the second term
corresponds to an in-coming wave. Similar procedures can be
applied to a negative $s(r)$ by assuming
$f_{n}(r)=B_{n}e^{-k_{0}\tau(r)}$, and the resultant first-order
approximation is given by
\begin{equation}
f_{n}(r)\sim\frac{B^{+}_{n}}{|s(r)|^{1/4}}\exp\left[-k_{0}\int_{r_{n}}^{r}\sqrt{|s(r')|}dr'\right].
\label{eq12}
\end{equation}
Here only the solution that is exponentially decaying in the $r$
direction is included. Furthermore, $s(r)\sim (r_{n}-r)$
approaches zero linearly in the vicinity of the turning point
$r_{n}$. The solution therefore can be approximated as
$\textrm{Ai}(-k_{0}^{2/3}s)$, with $\textrm{Ai}$ being the Airy
function \cite{chew}. When $k_{0}$ is large enough, we can
asymptotically match Equation (\ref{eq11}) and (\ref{eq12}) around
the turning point, and finally achieve $A^{+}_{n}=iA^{-}_{n}$. As
a direct result, in the region where $s(r)$ is positive, we have
\begin{equation}
f_{n}(r)\sim\frac{A_{n}}{s(r)^{1/4}}\exp\left[ik_{0}\int_{r_{n}}^{r}\sqrt{s(r')}dr'\right]+\frac{A_{n}}{s(r)^{1/4}}\exp\left[-i\frac{\pi}{2}-ik_{0}\int_{r_{n}}^{r}\sqrt{s(r')}dr'\right],
\label{eq13}
\end{equation}
which implies that an out-going wave will be totally reflected
around the turning point $r_{n}$, accompanied with a phase
variation of $\pi/2$.

\begin{figure}
\epsfxsize=210pt \epsfbox{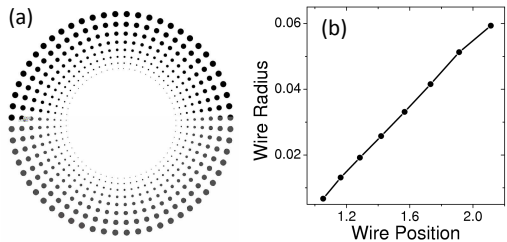} \vspace*{-6.0cm}
\caption{(a) A schematic of the design. (b) The radius of the
metallic wire as a function of its position. Here the permittivity
of the metal is $\epsilon_{m}=-0.6$.} \label{fig3}
\end{figure}

We now assume Equation (\ref{eq13}) can be extended to the region
where $r$ is slightly smaller than $1$. Additionally, its phase
factor can be approximated as
\begin{equation}
k_{0}\int_{r_{n}}^{r}\sqrt{s(r')}dr'=k_{0}\int_{r_{n}}^{r}\sqrt{s_{0}}dr'+k_{0}\int_{r_{n}}^{1}\left[\sqrt{s(r')}-\sqrt{s_{0}}\right]dr'\approx
k_{0}r+\gamma_{n}, \label{eq14}
\end{equation}
where $\gamma_{n}$ does not depend on $r$, and
\begin{equation}
k^{2}_{0}s_{0}=k_{0}^{2}-\frac{n^{2}-1/4}{r^{2}} \label{eq15}
\end{equation}
being the coefficient shown in Equation (\ref{eq5}). In the
vicinity of $r=1$, $f_{n}(r)$ can be then rewrite as
\begin{equation}
f_{n}(r)\sim\frac{A_{n}}{s_{0}^{1/4}}\left[e^{i(k_{0}r+\gamma_{n})}+e^{-i(k_{0}r+\gamma_{n}+\pi/2)}\right].
\label{eq16}
\end{equation}
Coincidentally, a source at position $\mathbf{r}_{0}$ with
$r_{0}<1$, \textit{i.e.} inside the Eaton lens, generates a
radiation such as
\begin{equation}
-\frac{i}{4}H_{0}^{(1)}(k_{0}|\mathbf{r}-\mathbf{r}_{0}|)=-\frac{i}{4}\sum_{-\infty}^{\infty}J_{n}(k_{0}r_{<})H_{n}^{(1)}(k_{0}r_{>})e^{in\phi},
\label{eq17}
\end{equation}
where $r_{<}$ is the smaller of $r$ and $r_{0}$, and $r_{>}$ is
the larger of $r$ and $r_{0}$ \cite{chew,jackson}. The total
magnetic field between $r_{0}$ and 1 is therefore given by
\begin{equation}
-\frac{i}{4}\sum_{-\infty}^{\infty}e^{in\phi}J_{n}(k_{0}r_{0})\left[H_{n}^{(1)}(k_{0}r)+C_{n}H_{n}^{(2)}(k_{0}r)\right],
\label{eq18}
\end{equation}
with $C_{n}$ representing the amplitude of the reflected $n$-th
order wave. When $k_{0}$ is large enough, the large argument
approximations of the Hankel functions and the Bessel function
lead to
\begin{equation}
J_{n}(k_{0}r_{0})\left[H_{n}^{(1)}(k_{0}r)+C_{n}H_{n}^{(2)}(k_{0}r)\right]\approx\frac{2}{k_{0}\pi\sqrt{rr_{0}}}
\cos(k_{0}r_{0}-\beta_{n})\left[e^{i(k_{0}r-\beta_{n})}+C_{n}e^{-i(k_{0}r-\beta_{n})}\right],
\label{eq19}
\end{equation}
with $\beta_{n}=(2n+1)\pi/4$. Again, we asymptotically match the
above equation with $f_{n}(r)/\sqrt{r}$ of Equation (\ref{eq16}),
and finally achieve
\begin{equation}
C_{n}\sim
e^{-i\left[(n+1)\pi+2\gamma_{n}\right]}=(-1)^{n+1}e^{-2i\gamma_{n}},
\:\: A_{n}\sim\frac{2s^{1/4}_{0}}{k_{0}\pi\sqrt{r_{0}}}
\cos(k_{0}r_{0}-\beta_{n}). \label{eq20}
\end{equation}

As mentioned at the beginning, an inside-out Eaton lens will
convert light rays emitted from a source at position
$\mathbf{r_{0}}$, with $r_{0}<1$, to the opposite location
$-\mathbf{r_{0}}$ and result in a perfect image. To extend this
property to waves, it is required that
\begin{equation}
C_{n}=(-1)^{n+1}e^{-i\phi}, \label{eq21}
\end{equation}
with $\phi$ being the phase difference between the source and
image. This relation can be easily obtained by time reversing the
source radiation process \cite{jackson}. Comparing Equation
(\ref{eq21}) with Equation (\ref{eq20}), we obtain the following
necessary condition for perfect imaging by an inside-out Eaton
lens: \textit{$e^{i2\gamma_{n}}$ must be constant and independent
of the mode order $n$}. The function $e^{i2\gamma_{n}}$ can then
be employed to partially evaluate the performance of a lens. For
instance, we calculate the phase factor $\gamma_{n}$ by using
Equation (\ref{eq14}), and the results are shown in Figure (2b).
Evidently, $\gamma_{n}$ are almost constant when $\lambda$ is
close to zero (the ray-optics region), while they vary strongly
for a considerable wavelength.

\begin{figure}
\epsfxsize=200pt \epsfbox{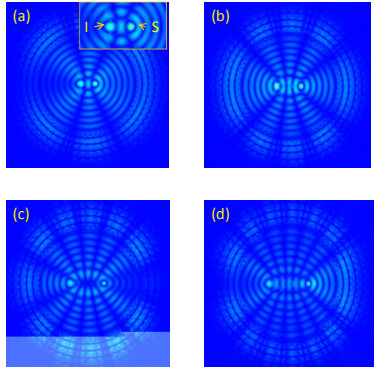} \vspace*{-4.0cm}
\caption{Full-wave simulations of the design, with the excitation
source placed at different locations. $r_{0}$ equals (a) 0.2, (b)
0.3, (c) 0.4, and (d) 0.5. The wavelength is fixed at 0.3, and the
permittivity of metal is $\epsilon_{m}=-0.6+0.01i$. The amplitude
of the magnetic field $|H|$ is plotted.} \label{fig4}
\end{figure}

In the following, we present a conceptual design consisting of a
vacuum (background) and ideal metallic wires (a realistic device
is currently under construction). First, the Maxwell-Garnett
formula is used to approximate the effective permittivity of the
composite (See Appendix B)
\begin{equation}
\epsilon_{e}=\frac{(1-f)+(1+f)\epsilon_{m}}{(1-f)\epsilon_{m}+(1+f)},
\label{eq22}
\end{equation}
with $f$ being the filling fraction of the wires and
$\epsilon_{m}$ being the permittivity of the ideal metal
\cite{jackson,yong}. Since $\epsilon_{e}$ should be equal to the
permittivity of the Eaton lens, $(2-r)/r$, the filling fraction is
then given by
\begin{equation}
f(r)=\left(r-1\right)\frac{1+\epsilon_{m}}{1-\epsilon_{m}}.
\label{eq23}
\end{equation}
Notice that the first-order surface mode will be excited when
$\epsilon_{m}=-1$ \cite{yong}. The designed lens, shown in Figure
(3a), consists of 8 layers of metallic wires with different radii.
The above equation is then used to determine the positions as well
as the radii of these wires. The thickness of the $p$-th layer
$a_{p+1}-a_{p}$ is set to be $a_{p}\pi/30$, with $a_{p}$ and
$a_{p+1}$ being the inner and outer radius of the layer,
respectively. Consequently we have
\begin{equation}
a_{p}=a_{1}(1+\pi/30)^{p-1}, \label{eq24}
\end{equation}
where $a_{1}$ is assumed to be 0.95. Furthermore the following
relation
\begin{equation}
\left(a_{p}-1\right)\frac{1+\epsilon_{m}}{1-\epsilon_{m}}\approx\frac{\pi
r^{2}_{p}}{a^{2}_{p}(\pi/30)^{2}}, \label{eq25}
\end{equation}
is employed to calculate the radius $r_{p}$ of the wire in the
$p$-th layer. In the current design, $\epsilon_{m}=-0.6+i\delta$
with $\delta$ being very small, which leads to the wire radii
shown in Figure (3b). Evidently, the wire radius increases rapidly
with the increasing of $r$, with a minimum of 0.007 and a maximum
of 0.06.

\begin{figure}
\epsfxsize=210pt \epsfbox{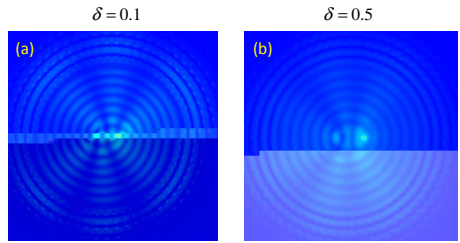} \vspace*{-6.0cm}
\caption{The effect of metallic loss $\delta$ on the lens. The
source is located at $r_{0}=0.2$. The wavelength $\lambda$ is set
to be 0.3 and $\textrm{Re}(\epsilon_{m})=-0.6$.} \label{fig5}
\end{figure}

We employ a finite-element full-wave Maxwell solver to verify the
above design \cite{comsol}, and plot the results with
$\delta=0.01$ in Figure (4). Notice that we always set
$\lambda=0.3$, a value much smaller than the lens size while big
enough to ensure the validity of Equation (\ref{eq22}). Clearly,
an image is always observed at the location opposite to the
source. To evaluate the quality of the image, we define
\begin{equation}
\eta=\left|\frac{H(-\mathbf{r}_{0})}{H(\mathbf{r}_{0})}\right|^{2},
\label{eq26}
\end{equation}
the ratio of the intensity of the image to that of the source. The
bigger the $\eta$, the better the lens is. In Figure (4), the
source location $r_{0}$ is gradually increased from 0.2 to 0.5,
with an increment of 0.1. The corresponding $\eta$ is found to be
0.21, 0.27, 0.15 and 0.15, respectively. Along the azimuthal
direction, we observe a few intensity minima, and the total number
of these minima depends on the source location $r_{0}$. This
phenomenon is very likely induced by the finite wavelength
$\lambda$. One direct consequence is that different modes radiated
from the source have quite different amplitudes, as given by
$J_{n}(k_{0}r_{0})$ of Equation (\ref{eq17}). For example, the
zeroth-order and third-order modes dominate the radiation when
$r_{0}=0.2$, while the ninth-order mode is the strongest one when
$r_{0}=0.5$. We further investigate the influence of the metallic
absorption in Figure (5), by setting $r_{0}=0.2$ and
$\lambda=0.3$. Two different $\delta$, 0.1 and 0.5, are
considered, and the corresponding intensity ratio $\eta$ is found
to be 0.11 and 0.06, respectively. Evidently, although the image
quality is degraded with the increasing of the metallic
absorption, the basic function of the Eaton lens, \textit{i.e.}
forming an image, still survives. We can partially interpret it
with the fact that all the higher-order modes, such as $n>25$, are
totally reflected around $r=1$. Only the low-order modes can
propagate into the lens and hence be absorbed by the metal.

To summarize, we have studied two-dimensional inside-out Eaton
lenses. The corresponding full wave equation is analytically
solved with the help of the WKB approximation. One necessary
condition for perfect imaging is further found, \textit{i.e},
$e^{i2\gamma_{n}}$ must be independent of the mode order $n$.
Furthermore, a general design procedure of the lens, based on
effective medium theory, is developed. We present one example
consisting of metal wires with different radii, and further verify
the design with a full wave Maxwell solver. Its dependence on
source location as well as metallic absorption is also
investigated.

This work was supported in part by the Penn State MRSEC under NSF
grant no. DMR 0213623.

\appendix
\numberwithin{equation}{section}

\section{A ray-optics theory of the Eaton lens}
A detailed description regarding the ray-optics theory of the
Eaton lens can be found in Reference \cite{leonhardt2}. We briefly
repeat it here for the reader's convenience.

In geometric optics, there are two different but equivalent ways
to describe the trajectory of a light ray. The first one is the
Newtonian Euler-Lagrange equation
\begin{equation}
\frac{d^{2}\mathbf{r}}{d\xi^{2}}=\frac{\nabla
n^{2}(\mathbf{r})}{2}, \label{ge1}
\end{equation}
where $n$ is the refractive index and the parameter $\xi$ is given
by $d\xi=dr/n$. We can interpret the above equation by using
Newton's law, $m\mathbf{a}=-\nabla U$, for a mechanical particle
with unit mass moving in "time" $\xi$ under the influence of
potential $U=-n^{2}/2+E$, with $E$ being an arbitrary constant.
The second way is based on Hamilton's equation
\begin{equation}
\frac{d\mathbf{r}}{dt}=\frac{c}{n}\frac{\mathbf{k}}{k},
\:\:\frac{d\mathbf{k}}{dt}=\frac{ck}{n^{2}}\nabla n(\mathbf{r}),
\label{ge2}
\end{equation}
with $\mathbf{k}$ being the wave vector and $c$ being the speed of
light in free space. Notice that by treating frequency
$\omega=ck/n$ as the Hamiltonian, the above equation resembles the
standard form of Hamilton's equation.

We can define an angular momentum as
\begin{equation}
\mathbf{L}=\mathbf{r}\times\frac{d\mathbf{r}}{d\xi}=\frac{n}{k}\mathbf{r}\times\mathbf{k},
\label{ge3}
\end{equation}
which leads to
\begin{equation}
\frac{d\mathbf{L}}{d\xi}=\mathbf{r}\times\frac{d^{2}\mathbf{r}}{d^{2}\xi}=\frac{1}{2}\mathbf{r}\times\nabla
n^{2}=\frac{dn^{2}}{dr}\frac{\mathbf{r}\times\mathbf{r}}{2r}=0,
\label{ge4}
\end{equation}
when the refractive-index profile $n(r)$ is spherically symmetric.
The above equation suggests that the angular momentum $\mathbf{L}$
is conserved. Hence, a family of light rays propagating in the
$xy$ plane at the beginning will always stay in the same plane.
This fact implies that a two-dimensional Eaton lens with similar
refractive-index profile $n(r)$ functions identically to the
three-dimensional one.

To solve the two-dimensional Newtonian Euler-Lagrange equation, it
is convenient to introduce the complex number $z=x+iy$, and
further reformulate the equation as
\begin{equation}
\frac{d^{2}z}{d\xi^{2}}=\frac{z}{2r}\frac{dn^{2}}{dr}=-\frac{1}{r^{3}}z=-\frac{1}{|z|^{3}}z,
\label{ge5}
\end{equation}
by substituting the refractive index of the Eaton lens $
n(r)=\sqrt{(2-r)/r}$. The solution, following Equation (6.13) and
(6.14) of Reference \cite{leonhardt2}, can be expressed as
\begin{equation}
z=e^{i\alpha}\left[\cos(2\xi')+i\sin\gamma\sin(2\xi')+\cos\gamma\right],\:\:
d\xi=2|z|d\xi',
\end{equation}
which describes displaced ellipses rotated by the angle $\alpha$.

\section{Maxwell-Garnett Formula}

Consider a two-component mixture composed of inclusions embedded
in an otherwise homogeneous matrix, where $\epsilon_{m}$ and
$\epsilon_{d}$ are their respective dielectric functions. The
average electric field $\langle\mathbf{E}\rangle$ over one unit
area surrounding the point $\mathbf{x}$ is defined as
\begin{equation}
\langle\mathbf{E}(\mathbf{x})\rangle=\frac{1}{A}\int_{A}\mathbf{E}(\mathbf{x}')d\mathbf{x}'=f\langle\mathbf{E}_{m}(\mathbf{x})\rangle+(1-f)\langle\mathbf{E}_{d}(\mathbf{x})\rangle,
\end{equation}
with $f$ being the volume fraction of inclusion. A similar
expression can be obtained for the average polarization
\begin{equation}
\langle\mathbf{P}(\mathbf{x})\rangle=f\langle\mathbf{P}_{m}(\mathbf{x})\rangle+(1-f)\langle\mathbf{P}_{d}(\mathbf{x})\rangle.
\end{equation}
We further assume that the following constitutive relations are
valid
\begin{equation}
\langle\mathbf{P}_{m}(\mathbf{x})\rangle=\epsilon_{0}(\epsilon_{m}-1)\langle\mathbf{E}_{m}(\mathbf{x})\rangle,
\:\:\:\langle\mathbf{P}_{d}(\mathbf{x})\rangle=\epsilon_{0}(\epsilon_{d}-1)\langle\mathbf{E}_{d}(\mathbf{x})\rangle,
\end{equation}
and the average permittivity tensor of the composite medium is
defined by
\begin{equation}
\langle\mathbf{P}(\mathbf{x})\rangle=\epsilon_{0}(\overline{\epsilon}_{e}-\mathbf{\overline{I}})\cdot\langle\mathbf{E}(\mathbf{x})\rangle.
\end{equation}
Combining the above equations we can obtain the effective
permittivity $\overline{\epsilon}_{e}$. Clearly the resultant
$\overline{\epsilon}_{e}$ depends on the relationship between
$\langle\mathbf{E}_{m}(\mathbf{x})\rangle$ and
$\langle\mathbf{E}_{d}(\mathbf{x})\rangle$ \cite{bohren}.

We now assume that the inclusion has the shape of a cylinder, and
its radius is far smaller than the wavelength so that its optical
properties can be well described by the electrostatic equation
\begin{equation}
\nabla\cdot(\epsilon(\mathbf{r})\phi)=0.
\end{equation}
By matching the boundary conditions we can prove that
$\phi_{m}/\phi_{0}=2\epsilon_{d}/(\epsilon_{d}+\epsilon_{m})$,
where $\phi_{m}$ is the total potential inside the cylinder when
the external electric field $-\nabla\phi_{0}$ is homogeneous. This
relation is further used to obtain the electric field
\cite{yong2}. It is finally found that the average permittivity is
scalar and can be expressed as
\begin{equation}
\epsilon_{e}=\epsilon_{d}\frac{(1-f)\epsilon_{d}+(1+f)\epsilon_{m}}{(1-f)\epsilon_{m}+(1+f)\epsilon_{d}},
\end{equation}
consistent with the Maxwell-Garnett dielectric function.
Equivalently we can express the filling fraction as
\begin{equation}
f(r)=\frac{(\epsilon_{e}-\epsilon_{d})(\epsilon_{m}+\epsilon_{d})}{(\epsilon_{e}+\epsilon_{d})(\epsilon_{m}-\epsilon_{d})}.
\end{equation}

\end{document}